\documentclass{aa}
\usepackage{graphicx}
\usepackage{txfonts}
\begin{document}
   \title{Initial-Final Mass Relationship for Stars of Different Metallicities}

%   \subtitle{I. Overviewing the $\kappa$-mechanism}

   \author{Xiangcun Meng,
          \inst{1,2}
          Xuefei Chen
          \inst{1}
          \and
          Zhanwen Han \inst{1}}
%\fnmsep\thanks{Just to show the usage
%          of the elements in the author field}

   \offprints{X. Meng}

   \institute{National Astronomical Observatories/Yunnan
Observatory, the Chinese Academy of Sciences, Kunming, 650011,
China\\
              \email{conson859@msn.com}
         \and
             Graduate School of the Chinese Academy of
Sciences\\
%             \email{c.ptolemy@hipparch.uheaven.space}
%             \thanks{The university of heaven temporarily does not
%                     accept e-mails}
             }

   \date{Received; accepted}

% \abstract{}{}{}{}{}
% 5 {} token are mandatory

  \abstract
   {The initial-final mass relationship (IFMR) for stars is important in many astrophysical fields, such as
   the evolution of galaxies, the properties of type Ia supernovae (SNe Ia) and the components of
   dark matter in the Galaxy.}
   {The purpose of this paper is to obtain the dependence of the IFMR on metallicity.}
   {Following Paczy\'{n}ski \& Zi\'{o}lkowski (1968) and Han et al. (1994),
   we assume that the envelope of an asymptotic giant branch (AGB) or
a first giant branch (FGB) star is lost when the binding energy of
the envelope is equal to zero ($\Delta W=0$) and the core mass of
the AGB star or the FGB star at the point ($\Delta W=0$) is taken
as the final mass. Using this assumption, we calculate the IFMRs
for stars of different metallicities.}
   {We find that the IFMR depends strongly on the metallicity, i.e. $Z=0.0001, 0.0003, 0.001, 0.004,
   0.01, 0.02, 0.03, 0.04, 0.05, 0.06, 0.08$ and $0.1$.
   From $Z=0.04$, the final mass of the stars with a given initial mass increases with increasing or decreasing
   metallicity. The difference of the final mass due to the metallicity may be up to 0.4 $M_{\odot}$.
   A linear fit of the initial-final mass relationship in NGC 2099 (M37)
   shows a potential evidence of the effect of metallicity on the IFMR.
   The IFMR for stars of $Z=0.02$ obtained in the paper
   matches well with
   those inferred observationally in the Galaxy. For
   $Z\geq 0.02$, helium WDs are obtained from the stars of $M_{\rm i}\leq 1.0 M_{\odot}$ and this result
   is upheld by the discovery of numerous low-mass WDs in NGC 6791 which is a metal-rich old open cluster.
   Using the IFMR for stars of $Z=0.02$ obtained in the
   paper, we have reproduced the mass distribution of DA WDs in Sloan DR4 except for some ultra-massive white
   dwarfs.}
   {The trend that the mean mass of WDs decreases with
   effective temperature may originate from the increase of the initial metallicities of stars.
   We briefly discuss the potential
   effects of the IFMR on SNe Ia and at the same time, predict that metal-rich low-mass
   stars may become under-massive white dwarfs.}

   \authorrunning{Meng, Chen \& Han}
   \titlerunning{The Initial-Final Mass Relationship for Different Metallicities }
   \maketitle{}
%
%________________________________________________________________

\section{Introduction}\label{sect:1}
White dwarfs (WDs) are the endpoint of the evolution of stars with
initial masses ranging from about 0.1 $M_{\odot}$ to about 8
$M_{\odot}$. The vast majority of stars in the Galaxy belong to
the mass range and over 97\% of the stars in the Galaxy will
eventually end up as WDs (Fontaine et al. \cite{FON01}). As a
result, WDs may give direct information about star formation
during the Galaxy's earliest epochs (Gates et al. \cite{GAT04}).
WDs show their importance in many fields. It has been about 50 yrs
since Schmidt (\cite{SCH59}) recognized the usefulness of WDs as
cosmochronometers. WDs may also be the dominating component of
dark matter in the Galaxy (Alcock et al. \cite{ALC99}; Chabrier
\cite{CHA99}). As a matter of fact, WDs are very important for
type Ia supernovae (SNe Ia) since it is believed that SNe Ia are
from the thermonuclear runaway of carbon-oxygen white dwarfs (CO
WD) (see the reviews by Hillebrandt \& Niemeyer (\cite{HN00}) and
Leibundgut (\cite{LEI00})). It is well known that the masses of
WDs are typically of the order of half that of the Sun, while
their radii are similar to that of a planet. However, a detailed
knowledge of WDs is still unclear in observation and theory
(Fontaine et al. \cite{FON01}; Moroni \& Straniero \cite{MOR02},
\cite{MOR07}). Among all the uncertainties of WDs, the correlation
between initial-final mass relationship (IFMR) and metallicitiy is
a very important one. It is well known that low metallicity leads
to a larger CO WD for a given initial mass with $Z\leq 0.02$
(Umeda et al. \cite{UME99a}). However, it is necessary to check
the cases of $Z>0.02$ since $Z\in[0.06-0.1]$ is possible for some
ultra-luminous galaxies (Roberts \& Hynes \cite{RH94};
Ruiz-Lapuente et al. \cite{RUI95}; Terlevich \& Forbes
\cite{TER02}).

The IFMR for stars over a large mass range (e.g. 0.8-8
$M_{\odot}$) is a powerful input to chemical evolution models of
galaxies (including enrichment in the interstellar medium) and
therefore can enhance our understanding about star formation
efficiencies in these systems (Ferrario et al. \cite{Fer05};
Kalirai et al. \cite{KAL07b}). The IFMR is also an important input
for modelling the luminosity functions of Galactic disk WDs and
the cooling sequences of halo clusters, which may directly yield
the age of the Galactic disk and halo components (Kalirai et al.
\cite{KAL07b}). Meanwhile, the IFMR represents the mass loss of a
star over its entire evolution and it is possible to get some
indications about the origin and evolution of hot gas in
elliptical galaxies from the IFMR (Mathews \cite{MAT90}). Since
Weidemann (\cite{WEI77}) showed the first comparison between
observations and theoretical predications of the IFMR, many
observations gave constraints on the relationship by studying the
properties of white dwarfs in open clusters or field white dwarfs,
especially during the last decade (Herwig \cite{HER95}; Reid
\cite{REI96}; Koester \& Reimers \cite{KR96}; Finley \& Koester
\cite{FK97}; Claver et al. \cite{CLA01}; Williams et al.
\cite{WIL04}; Ferrario et al. \cite{Fer05}; Williams et al.
\cite{WIL06}). Some empirical relations have also been suggested
based on different observations (Weidemann \cite{WK83},
\cite{WEI00}; Ferrario et al. \cite{Fer05}; Williams et al.
\cite{WIL06}, Dobbie et al. \cite{DOB06b}). At the same time,
numerous sets of stellar evolution models have been calculated to
study the relationship in theory and to give the IFMRs for stars
of different metallicity (Han et al. \cite{HAN94}; Girardi et al.
\cite{GIR00}). Although great progress has been made in
observation and theory, there still exist many uncertainties, i.e.
what is the origin of the intrinsic scatter of WD mass or whether
there is any dependence of the IFMR on the metallicity (Kalirai et
al. \cite{KAL05}; Williams \cite{WIL06}). The correlation between
the IFMR and the metallicity has been established in theory for
many years (Han et al. \cite{HAN94}; Girardi et al. \cite{GIR00}),
but the evidence of the dependence of the IFMR on the metallicity
was not found until 2005 (Kalirai et al. \cite{KAL05}). Many
theoretical calculations showed that the core mass at the first
thermal pulse (TP) in the asymptotic giant branch (AGB) may be
taken as the final mass (see the review by Weidemann
\cite{WEI00}). However, these theoretical studies only focused on
some special metallicities and it is also difficult for some
stellar evolution codes to determine which is the first thermal
pulse. In the paper, we use a simple but robust method to
systemically study the IFMR over a wide metallicity range, i.e.
$0.0001\leq Z\leq 0.1$. We also briefly discuss the potential
effect of the IFMR on SNe Ia.

In section \ref{sect:2}, we describe our model and physical
inputs. We give the results in section \ref{sect:3} and show
discussions and conclusions in section \ref{sect:4}.

   \begin{figure*}
   \centering
%   \vspace{2mm}
%   \begin{center}
   %%%\includegraphics{empty.eps}
   %%%\includegraphics{empty.eps}
   \includegraphics[width=150mm,height=160mm,angle=270.0]{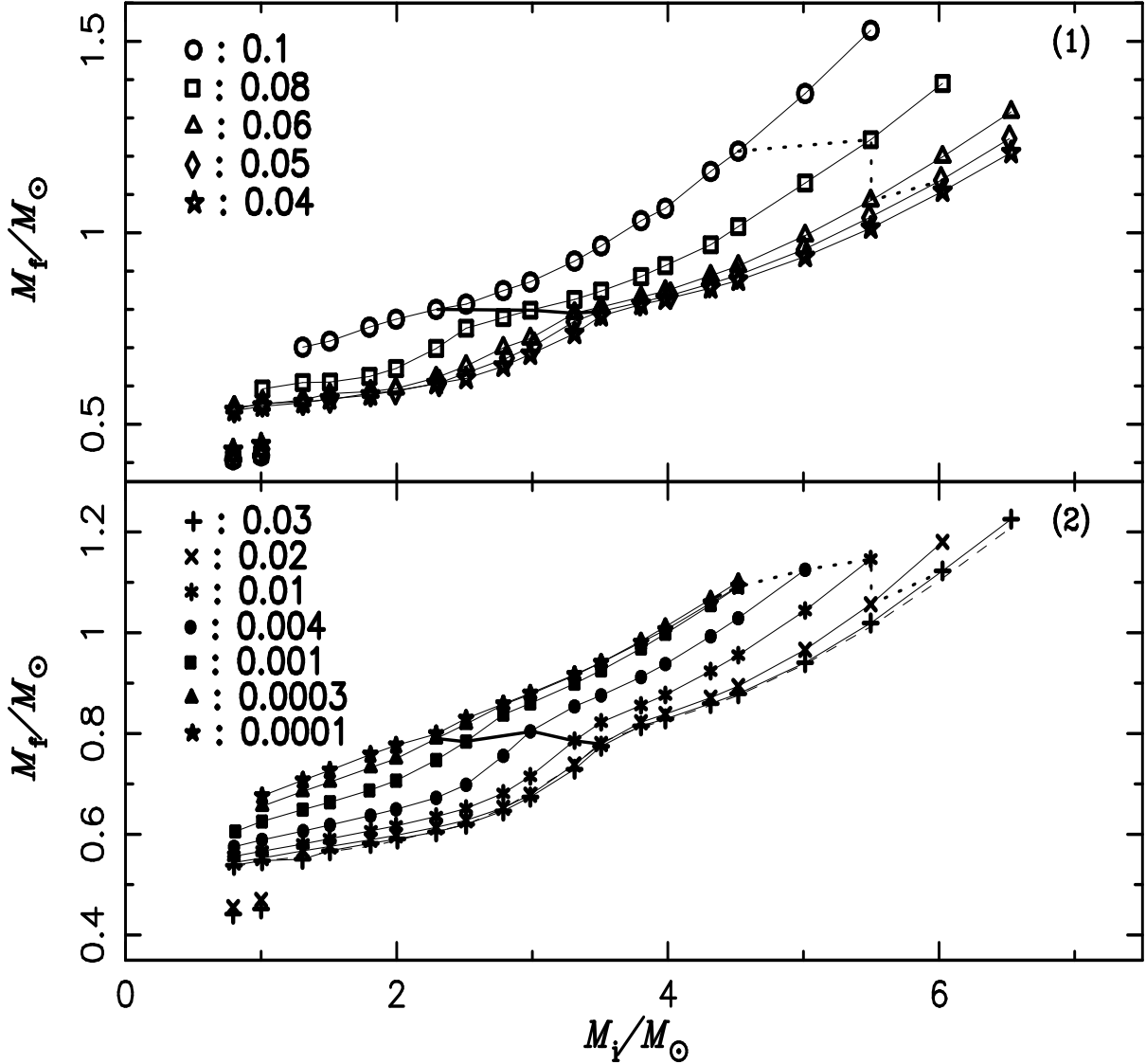}
%   \hspace{3mm}
%   \psfig {figure=mix.ps,width=80mm,height=60mm,angle=270.0}
%   \parbox{180mm}{{\vspace{2mm} }}
   \caption{Initial-final mass relationship (IFMR) for stars of different metallicities. The points of $M_{\rm f}< 0.5 M_{\odot}$
   represent helium WDs. The thick
   solid lines are the mass threshold for the second dredge-up (see the text for details) and the thick
   dotted ones show the transition between carbon-oxygen white dwarfs and
   oxygen-neon-magnesium ones in our models.
   The dashed line in the lower panel shows the IFMR for stars of $Z=0.04$ as a comparison. }
              \label{Fig1}%
    \end{figure*}

%__________________________________________________________________

\section{Model and Physical Inputs}\label{sect:2}

%                                     Two column figure (place early!)
%______________________________________________ Gamma_1 (lg rho, lg e)
\subsection{the model}\label{subs:2.1}
\begin{figure}
   \centering
%   \vspace{10mm}
%   \begin{center}
   %%%\includegraphics{empty.eps}
   %%%\includegraphics{empty.eps}
   \includegraphics[width=85mm,height=80mm,angle=270.0]{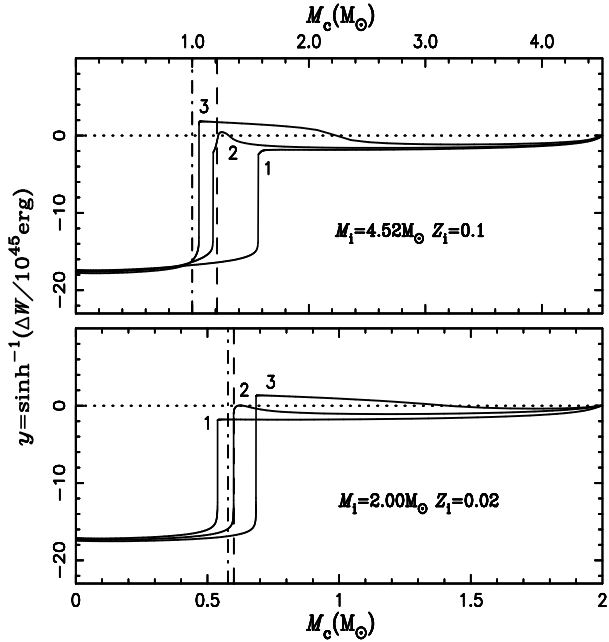}
\hspace{10mm}
   \parbox{100mm}{{\vspace{6mm} }}
   \caption{The change of binding energy $\Delta W$ with the different choices of core mass,
   $M_{\rm c}$, (equation (1))
   at three evolutionary stages for two AGB stars. bottom panel is for $M_{\rm i}=2.00 M_{\odot}, Z_{\rm i}=0.02$
   and the top panel is for $M_{\rm i}=4.52 M_{\odot}, Z_{\rm i}=0.1$.
   The numbers present time sequence.
   The dependent variable, $y={\rm sinh^{\rm -1}}(\Delta W/10^{\rm
   45}\rm erg)$, keeps the sign of $\Delta W/10^{\rm
   45}\rm erg$ and is a logarithmic function of $\Delta W/10^{\rm
   45}\rm erg$ when its absolute value is large, or a linear function when its absolute value is small.
   The dashed lines show the boundaries of the cores of the models defined in this paper,
   while the dot-dashed
   ones show the positions of helium burning shells.}
              \label{Fig2}%
    \end{figure}

When a star is in the AGB stage, its envelope may be blown off if
the binding energy (BE) of the envelope changes from negative to
positive (Paczy\'{n}ski \& Zi\'{o}lkowski \cite{PZ68}). It is
clearly shown in Fig. 1 of Han et al. (\cite{HAN94}) that the BE
of an AGB star changes from negative to positive through the AGB
evolution (also Fig. \ref{Fig2} in this paper). The BE of the
envelope can be calculated by
  \begin{equation}
 \Delta W=\int_{M_{\rm c}}^{M_{\rm s}}(-\frac{Gm}{r}+U){\rm d}m,
  \end{equation}
where $M_{\rm c}$ is the core mass, $M_{\rm s}$ is the surface
value of the mass coordinate $m$, and $U$ is the internal energy
of thermodynamics (including terms due to ionization of H and
dissociation of $\rm H_{\rm 2}$, as well as the basic
$\frac{3}{2}\Re T/\mu$ for a perfect gas). We assume that the
envelope of a star is lost if $\Delta W=0$ and the core mass at
this point is the final WD mass. The virtue of this method is that
the mechanism of mass loss need not be considered. Han et al.
(\cite{HAN94}) used this assumption to get the IFMR for stars of
$Z=0.02$ and $Z=0.001$. The binary population synthesis (BPS)
incorporating their IFMRs well reproduced the mass distribution of
planetary nebula nuclei (PNN). Their result indicated that this
method is robust to calculate the final mass for a star of given
initial mass and metallicity (see Han et al. \cite{HAN94} for
details).

In principle, the envelope has enough energy to escape to infinity
when $\Delta W>0$. As mentioned in Han et al. (\cite{HAN94}),
however, it is not clear whether the instant envelope ejection
occurs or not. Radiation might take away some of the energy which
is required to go into outward motion in order to achieve the
envelope ejection, but it is possible that a series of
oscillations on a dynamical time-scale with amplitude growing
occur until $\Delta W$ is large enough to energize envelope
ejection (Han et al. \cite{HAN94}). The discussion above means
that $\Delta W\geq0$ may only be a necessary condition to eject
the envelope of an AGB star. Our assumption implies that a
superwind starts at or after $\Delta W=0$. Fortunately, the change
of the core mass during the superwind phase is negligible at low
masses and relatively modest at high masses (Han et al.
\cite{HAN94}).

For saving cpu time, we simply treated the average evolution of
thermally pulsing AGB models, i.e. skipping thermal pulses by
taking a longer time-step. We give some discussions on the
influence of TPs and mass loss in subsection \ref{subs:4.1}.

\subsection{Physical Inputs}\label{subs:2.2}
We use the stellar evolution code of Eggleton (\cite{EGG71},
\cite{EGG72}, \cite{EGG73}), which has been updated with the
latest input physics over the last three decades (Han et al.
\cite{HAN94}; Pols et al. \cite{POL95}, \cite{POL98}). We set the
ratio of mixing length to local pressure scale height,
$\alpha=l/H_{\rm p}$, to 2.0, and set the convective overshooting
parameter, $\delta_{\rm OV}$, to 0.12 (Pols et al. \cite{POL97};
Schr\"{o}der et al. \cite{SCH97}), which roughly corresponds to an
overshooting length of $0.25 H_{\rm P}$. The range of metallicity
is from 0.0001 to 0.1, i.e. 0.0001, 0.0003, 0.001, 0.004, 0.01,
0.02, 0.03, 0.04, 0.05, 0.06, 0.08, 0.1. The opacity tables for
these metallicties are compiled by Chen \& Tout (\cite{CHE07})
from Iglesias \& Rogers (\cite{IR96}) and Alexander \& Ferguson
(\cite{AF94}). For a given $Z$, the initial hydrogen mass fraction
is assumed by

 \begin{equation}
 X=0.76-3.0Z,
  \end{equation}
(Pols et al. \cite{POL98}), and then the helium mass fraction is
$Y=1-X-Z$. Based on the correlation among $X$, $Y$ and $Z$ used
here, Pols et al. (\cite{POL98}) accurately reproduced the
color-magnitude diagrams (CMD) of some clusters.

The mass loss might affect the IFMR. Han et al. (\cite{HAN94})
have shown that the steady stellar wind based on the observational
relation given by Judge \& Stencel (\cite{JS91}) can not
significantly affect the IFMR. We also test the effect of Reimers'
wind (\cite{REI75}) with $\eta=1/4$ on the IFMR of $Z=0.02$. A
similar result to that in Han et al. (\cite{HAN94}) is obtained.
Buzzoni et al. (\cite{BUZ06}) also noted that Reimers' wind
parametrization poorly reproduce the relative number of planetary
nebulae (PNe) in late-type galaxies, while a better fit for the
relative number of PNe is obtained using the empirical IFMR in
Weidemann (\cite{WEI00}). This might imply that the Reimers' wind
is a unreasonable mechanism to calculate the IFMR. We therefore do
not include any wind mass loss in our calculation for any of the
metallicities.

\section{Results}\label{sect:3}
\subsection{the initial-final mass relationship for stars of different metallicities}\label{subs:3.1}
Fig. \ref{Fig1} shows the IFMRs for stars of different
metallicities. The IFMRs in the paper may be approximated by two
parabola segments, which is similar to that of Han et al.
(\cite{HAN94}) for $Z=0.02$ and $Z=0.001$. The fitted formulae of
the IFMRs for stars of different metallicities are shown in
Appendix A.

Han et al. (\cite{HAN94}) showed that the BE of the envelope of
Pop I low-mass stars ($M_{\rm i}\leq 1.0$) may get the point of
$\Delta W=0$ on the first giant branch (FGB) and helium WDs may be
obtained. This phenomenon was used to explain the existence of low
mass WD in Stein 2051B and 40 EriB. We get a similar result to
that of Han et al. (\cite{HAN94}) for those stars with $Z\geq
0.02$ and  $M_{\rm i}\leq 1.0 M_{\odot}$ (See those $M_{\rm f}<0.5
M_{\odot}$ in Fig. \ref{Fig1}). The final masses of the He WDs for
a given initial mass slightly depend on the metallicity. For a
given initial mass, a larger He WD is obtained for a small
metallicity (see in Table \ref{Tab:1}). This is because for a
perfect gas, mean molecular weight, $\mu$, increases with
metallicity according to equation (2) (Kippenhahn \& Weigert
\cite{KW90}) and the thermal energy therefore decreases with the
metallicity. Though the final radii of the low-mass stars when
$\Delta W=0$ in FGB slightly increases with the metallicity (see
in Table \ref{Tab:2}), it does not significantly affect the
gravitational energy  of the envelope. So, the core mass
dominating the gravitational energy of the envelope decreases with
the metallicity to counteract the decrease of the thermal energy.

According to the above finding, there are probably several
under-massive WDs ($M_{\rm WD}< 0.5 M_{\odot}$) in metal-rich old
clusters. Observationally, Kalirai et al. (\cite{KAL07a}) recently
found that most WDs are under-massive and the mean mass of the WDs
is $<M>=0.43\pm0.06M_{\odot}$ in NGC 6791, which is one of the
oldest (8-12Gyr) and most metal-rich ([Fe/H]=+0.4$\pm$0.1) open
clusters in the Galaxy (Peterson \& Green \cite{PET98}; Chaboyer
et al. \cite{CHAB99}; Stetson et al. \cite{STE03}; Gratton et al.
\cite{GRA06}; Origlia et al. \cite{ORI06}). According to the high
metallicity and the old age of NGC 6791, the stars with $M_{\rm
i}\sim 1.0 M_{\rm \odot}$ in NGC 6791 likely have left the main
sequence and moved into the FGB phase (the life of stars of
$M_{\rm i}=1.0 M_{\odot}$ and $Z=0.04-0.06$ on the main sequence
is about 7.5-9 Gyr, Chen \& Tout \cite{CHE07}). Kalirai et al.
(\cite{KAL07a}) suggested that the under-massive stars are helium
WDs which have experienced enough mass loss on the FGB to avoid
the helium flash, which naturally resolves the age discrepancy of
the cluster from white dwarf cooling theory (2.4 Gyr) and from
main-sequence turnoff (8 - 12 Gyr) (Bedin et al. \cite{BED05}).

It is obvious from Fig \ref{Fig1} that there is a dependence of
IFMR on the metallicity. The IFMR gets to its lower limit at
$Z=0.04$, i.e. the final mass is smallest for $Z=0.04$ for a given
initial mass. It is well known that for stars with $Z\leq 0.02$,
the effect of decreasing metallicity is similar to that of
increasing mass, i.e. the evolution track of a low-mass star with
a low metallicity is similar to that of a larger one with a higher
metallicty (Umeda et al. \cite{UME99a}). The result of Chen \&
Tout (\cite{CHE07}) reconfirmed the fact and suggested an opposite
trend when $Z\geq 0.04$, i.e. for the stars with a given initial
mass, the age decreases with the metallicity and the luminosity
increases with the metallicity (see Fig. 2 and Table 1 in that
paper). The phenomenon that the effect of the metallicity on the
evolution of a star is not monotonic is derived from the
correlation among $X$, $Y$ and $Z$ used in this paper. The three
components all contribute to opacity which dominates the evolution
of a star with a certain mass. As is well known, the opacity
increases with $X$ or $Z$ and decreases with $Y$. The effect of
increasing opacity is similar to that of decreasing mass. For
$Z<0.04$, the change of $Z$ dominates the change of the opacity,
while the changes of $Y$ and $X$ dominate the change of the
opacity for $Z\geq 0.04$.

According to our calculation, the metallicities will result in a
scatter of the final masses. For a certain mass, the scatter may
be up to 0.4 $M_{\odot}$, which depends on the differences of
metallicity and initial stellar mass. Williams (\cite{WIL06})
compared the IFMR for stars in NGC 2099 (M37, Z=0.01 (Kalirai et
al. \cite{KAL05})) with that in other two clusters ( Hyades and
Praesepe, Z=0.02 (Perryman et al. \cite{PER98}; Claver et al.
\cite{CLA01})) and claimed that ``any metallicity dependence of
the IFMR for $M_{\rm i}\approx 3 M_{\odot}$ must be smaller than
$\Delta M_{\rm f}\approx 0.05 M_{\odot}$". The difference of the
final mass for $M_{\rm i}\approx 3 M_{\odot}$ between $Z=0.02$ and
$Z=0.01$ in this paper is about $0.04 M_{\odot}$, which is
consistent with the result of Williams (\cite{WIL06}).

We assume that the remnant is a CO WD if carbon and oxygen have
not been ignited in a star when $\Delta W=0$ (see the thick dotted
lines in Fig. \ref{Fig1}). In Table \ref{Tab:3}, we show the
maximum masses of the CO WDs, $M_{\rm max}^{\rm CO}$, for
different metallicities. Note that the values of the $M_{\rm
max}^{\rm CO}$ probably have some differences from the real ones
because of the grid density of the models in this paper. In our
study, the $M_{\rm max}^{\rm CO}$ are $1.2431 M_{\odot}$ and
$1.2132 M_{\odot}$ for $Z=0.08$ and $Z=0.1$, respectively.
Generally, the mass of a CO WD is less than 1.20 $M_{\odot}$,
otherwise carbon will be ignited in the region with the highest
temperature and the final remnant after the ejection of its
envelope is an oxygen-neon-magnesium (ONeMg) WD. The overestimated
values in our models for $Z=0.08$ and $Z=0.1$ are mainly from the
determination of the core mass $M_{\rm c}$. As discussed by Han et
al. (\cite{HAN94}), it is difficult to determine where the
boundary of the core is. As an examination, we choose different
core masses $M_{\rm c}$ from the center to the surface and
calculated the BE $\Delta W$ as a function of $M_{\rm c}$. Three
evolutionary stages, i.e. before, at and after $\Delta W=0$, are
selected to check the evolution of $\Delta W=0$ for two AGB stars.
The results are shown in Fig. \ref{Fig2}. In the figure, we see
that there are three portions: (i) an inner region where $\Delta
W$ increases slowly; (ii) a zone where $\Delta W$ increases
sharply, the zone including the H-burning shell; and (iii) an
outer portion where $\Delta W$ varies slowly with $M_{\rm c}$. We
take the value near but outside the transition between (ii) and
(iii) as $M_{\rm c}$ (the dashed lines in the figure) to be sure
that $\Delta W$ is not very sensitive to $M_{\rm c}$. This choice
is good enough for stars with a thin transition between (ii) and
(iii), i.e. both $M_{\rm i}$ and $Z$ are not very large such as
$M_{\rm i}=2.00 M_{\odot}$ and $Z=0.02$ in the figure. However,
for those with large initial mass and high metallicity, the
transition between (ii) and (iii) is thick (see the model of
$M_{\rm i}=4.52M_{\odot}, Z_{\rm i}=0.1$ in Fig. \ref{Fig2}) and
it is difficult to choose the boundary of the core. This may
result in an uncertainty of the final mass by $0.03 M_{\odot}$ for
the model of $M_{\rm i}=4.52M_{\odot}, Z_{\rm i}=0.1$. Actually,
for most of the models in the paper, our choice of $M_{\rm c}$ is
not a serious problem. The maximum uncertainty of the final masses
derived from the choice of $M_{\rm c}$ is about $0.04 M_{\odot}$,
which is from the model of $M_{\rm i}=5.50M_{\odot}$ and $Z_{\rm
i}=0.1$.

The value of $M_{\rm max}^{\rm CO}$ might also be overestimated if
the star reaches $\Delta W=0$ again after the ejection of H-rich
envelope. There is still a helium envelope {around the CO core
after a star loses its hydrogen-rich envelope and the helium
envelope is thicker with $M_{\rm i}$ and $Z$. For example, the
mass of the helium envelope after the loss of the hydrogen-rich
envelope may be as large as 0.2 $M_{\odot}$ for a star of $M_{\rm
i}=4.52 M_{\odot}$ with $Z=0.1$ (see in Fig \ref{Fig2}, the
dot-dashed lines show the positions of the helium burning shells).
Then, the star becomes a helium red giant. With the expansion of
the helium envelope, the envelope might get the point where
$\Delta W=0$ and a part of the envelope might be lost again, which
may lead to an overestimate of $M_{\rm max}^{\rm CO}$ (Han et al.
\cite{HAN94}).

In Fig. \ref{Fig2}, we see that, although the shape of $\Delta W$
with $M_{\rm c}$ has not changed during the AGB phase, the time
sequence is different for the two stars, i.e. the region of the
sharp increase of $\Delta W$ moves outward with time for the
low-mass star, while it is opposite for the high-mass one. This
difference comes from whether the second dredge-up occurs or not
in a star. Generally, for Pop I stars (Z = 0.02), the second
dredge-up occurs in early-AGB (EAGB) stars defined by Iben
(\cite{IBEN83}) if $M_{\rm i}>3.5M_{\odot}$ (Busso et al.
\cite{BGW99}), which reduces the mass of H-exhausted core (see
Fig. 6 of Meng et al. \cite{MCTH06}). $\Delta W$ reaches zero just
at the process of core decreasing as shown in the upper panel of
Fig. \ref{Fig2}. For the stars with $M_{\rm i}<3.5M_{\odot}$, the
second dredge-up does not occur and the core mass always increases
with time as shown in the bottom panel of Fig. \ref{Fig2}. When a
star has a mass around $3.5 M_{\odot}$, the core mass is almost
constant with the expansion of its envelope until $\Delta W=0$
(see Fig. 6 of Meng et al. \cite{MCTH06}). The mass threshold for
the second dredge-up changes with metallicity $Z$. We presented
the mass threshold from our models for various $Z$ in Table
\ref{Tab:4} (see the thick solid line in Fig.
\ref{Fig1}).$\footnote{Note that the mass threshold may have an
uncertainty of $0.15 M_{\odot}$ because of the grid density of the
models in this paper. The mass interval is about $0.2-0.3
M_{\odot}$ for our models.}$ From Table \ref{Tab:4}, we see that,
since the stellar evolution is non-monotonic with $Z$,
$\footnote{The non-monotonic effect of metallicity on stellar
evolution is from equation (1). Please see the fourth paragraph in
subsection \ref{subs:3.1}.}$ the mass threshold is also
non-monotonic with $Z$.

We notice that the final mass of the star of $M_{\rm i}=5.5
M_{\odot}$ with $Z=0.1$ is larger than 1.5 $M_{\odot}$. It is
likely that this remnant is not a white dwarf. When the star first
arrives at the point of $\Delta W=0$, carbon and oxygen have been
ignited and the CO core will become an ONeMg core. As described in
the paragraph above, the star will become a helium giant star
after the ejection of its hydrogen-rich envelope. If the helium
envelope can not get to the point of $\Delta W=0$, the mass of the
ONeMg core might arrive at the Chandrasekhar mass limit after
about 1000 - 10000 yrs and the star might explode as a type Ib
supernova. Otherwise, an ONeMg WD would be its final fate.

For the cases of extremely low and high metallicities, the
situation of $\Delta W=0$ can not be met at the low-mass end and
we can not obtain the final mass through AGB stars (see Fig
\ref{Fig1}). This is because for the stars with these extreme
condition, the final mass is too large with respect to the initial
mass and the envelope is too thin. For example, the mass of the
core through AGB is larger than 0.6 $M_{\odot}$ for a star of
$M_{\rm i}=0.8 M_{\rm \odot}$ with $Z=0.0001$, and the situation
of $\Delta W = 0$ is not obtained before Eggleton's stellar
evolution code can not work. This may imply that if the low-mass
stars with extremely low and high metallicities experienced the
AGB stage and became CO WDs finally, they might have a different
mechanism to lose their envelopes from those with middle
metallicities.

   \begin{figure*}
   \centering
%   \vspace{2mm}
%   \begin{center}
   %%%\includegraphics{empty.eps}
   %%%\includegraphics{empty.eps}
   \includegraphics[width=150mm,height=160mm,angle=270.0]{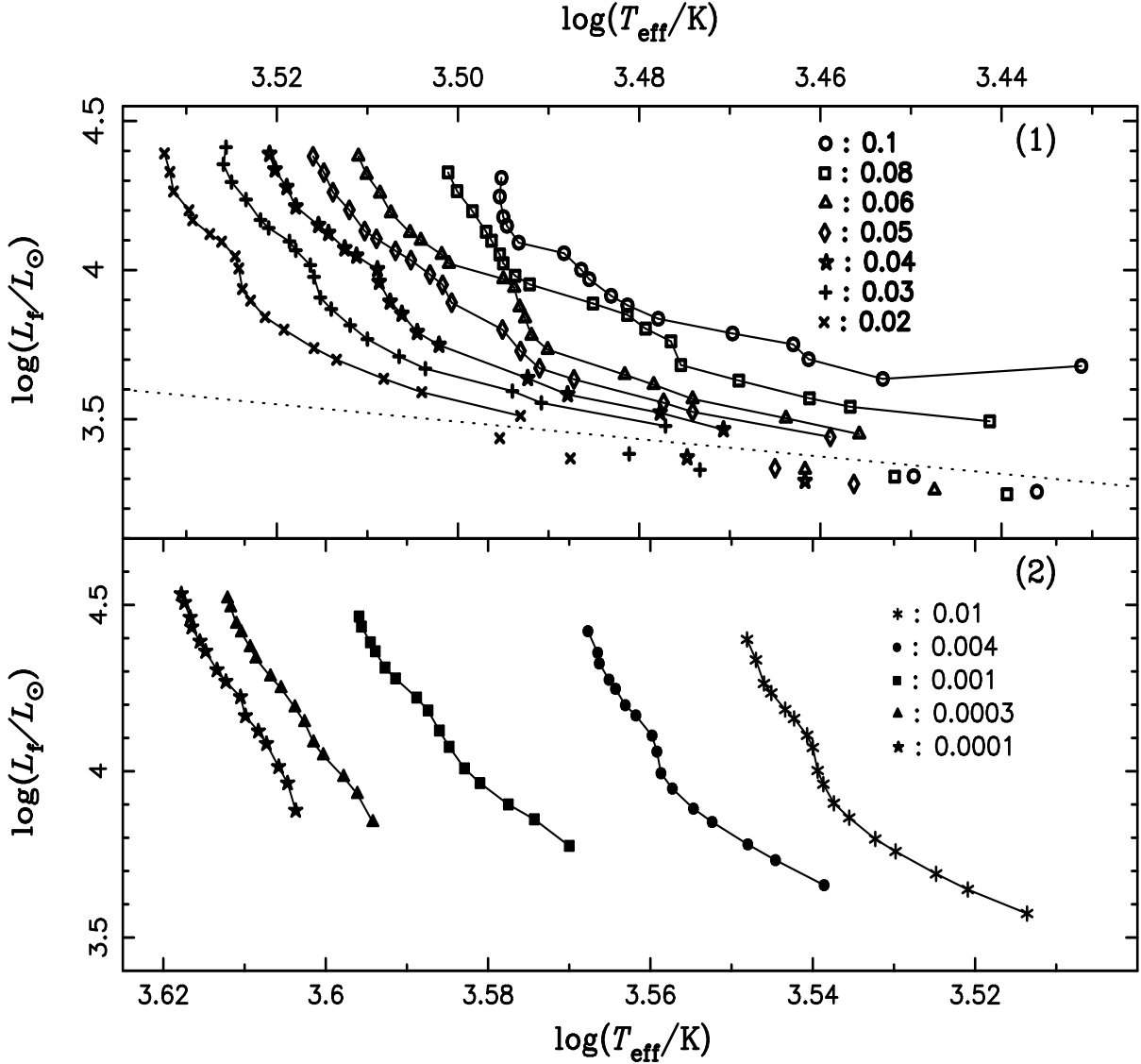}
%   \hspace{3mm}
%   \psfig {figure=mix.ps,width=80mm,height=60mm,angle=270.0}
%   \parbox{180mm}{{\vspace{2mm} }}
   \caption{The final positions ($\Delta W=0$) of FGB/AGB in Hertzsprung-Russel diagram (HRD)
   for different metallicities. In the upper panel, the final positions of AGB stars locate above the dotted line,
   while the final positions of FGB stars locate under it. The upper panel shows models whose metallicity is
   high enough to permit the low-mass stars to get to the point of $\Delta W=0$ on the FGB. Note that
   the scale of the
   abscissa in the upper panel is different from that in the lower panel.}
              \label{Fig3}%
    \end{figure*}

\subsection{Hertzsprung-Russel diagram}\label{subs:3.2}
The final positions ($\Delta W=0$) of the stars before envelope
loss with the different metallicities in a Hertzsprung-Russel
diagram (HRD) are shown in Fig. \ref{Fig3}. The dotted line gives
the division of the FGB and AGB stars (the upper populations are
AGB stars, while the lower ones are FGB stars). The effective
temperature strongly depends on the metallicity since all the
stars are near the Hayashi line which is determined by
metallicity.

\begin{table*}[]
\caption[]{The masses (in $M_{\odot}$) of helium WDs for different
metallicities (in $Z_{\odot}$) (Row 1) and different initial
masses (Column 1)(in $M_{\odot}$).} \label{Tab:1}
\begin{center}
\begin{tabular}{cccccccc}
\hline\noalign{\smallskip}%\scriptsize
$M_{\rm i}$& 1.0    & 1.5    & 2.0    & 2.5    & 3.0    & 4.0    & 5.0  \\
\hline\noalign{\smallskip}
 0.79      & 0.4554 & 0.4415 & 0.4311 & 0.4257 & 0.4185 & 0.4108 & 0.4074\\
 1.00      & 0.4697 & 0.4518 & 0.4463 & 0.4354 & 0.4318 & 0.4213 & 0.4176\\
\noalign{\smallskip}\hline
\end{tabular}\end{center}
\end{table*}

\begin{table*}[]
\caption[]{The final radius (in $\log R_{\rm f}/R_{\odot}$) when
$\Delta W=0$ in first giant branch for different metallicities (in
$Z_{\odot}$) (Row 1) and different initial masses (Column 1)(in
$M_{\odot}$).} \label{Tab:2}
\begin{center}
\begin{tabular}{cccccccc}
\hline\noalign{\smallskip}%\scriptsize
$M_{\rm i}$& 1.0    & 1.5    & 2.0    & 2.5    & 3.0    & 4.0    & 5.0  \\
\hline\noalign{\smallskip}
 0.79      & 2.2323 & 2.2419 & 2.2462 & 2.2523 & 2.2596 & 2.2688 & 2.2796\\
 1.00      & 2.2507 & 2.2531 & 2.2595 & 2.2610 & 2.2658 & 2.2734 & 2.2784\\
\noalign{\smallskip}\hline
\end{tabular}\end{center}
\end{table*}

\begin{table*}[]
\caption[]{The maximum masses (Row 3) of CO WDs for different
metallicities (Row 1). The initial masses to get the final maximum
CO WD are shown in Row 2.} \label{Tab:3}
\begin{center}
\begin{tabular}{ccccccc}
\hline\noalign{\smallskip}%\scriptsize
$Z/Z_{\odot}          $ &0.005 &0.015 &0.05  &0.2   &0.5   &1.0   \\
\hline\noalign{\smallskip}
$M_{\rm i}/M_{\odot}  $ &4.52  &4.32  &4.52  &5.01  &5.50  &5.50  \\
$M_{\rm max}^{\rm CO}/M_{\odot}$ &1.0918&1.0667&1.0912&1.1251&1.1461&1.0566\\
\hline\noalign{\smallskip}%\scriptsize
$Z/Z_{\odot}          $ &1.5   &2.0   &2.5   &3.0   &4.0   &5.0  \\
\hline\noalign{\smallskip}
$M_{\rm i}/M_{\odot}  $ &6.03  &6.03  &6.03  &5.50  &5.50  &4.52 \\
$M_{\rm max}^{\rm CO}/M_{\odot}$ &1.1228&1.1081&1.1370&1.0842&1.2431&1.2132\\
\hline\noalign{\smallskip} %\noalign{\smallskip}\hline
\end{tabular}\end{center}
\end{table*}

\begin{table*}[] \caption[]{Mass threshold (Row 2, in solar mass) for
different metallicities (Row 1, in solar metallicity). If a star
has an initial mass larger than the mass threshold for a certain
metallicity, the second dredge-up occurs in the star when helium
is exhausted in the center.} \label{Tab:4}
\begin{center}
\begin{tabular}{ccccccc}
\hline\noalign{\smallskip}%\scriptsize
$Z/Z_{\odot}          $ &0.005 &0.015 &0.05  &0.2   &0.5   &1.0   \\
\hline\noalign{\smallskip}
$M_{\rm i}/M_{\odot}  $ &2.30  &2.30  &2.50  &3.00  &3.30  &3.50  \\
\hline\noalign{\smallskip}%\scriptsize
$Z/Z_{\odot}          $ &1.5   &2.0   &2.5   &3.0   &4.0   &5.0  \\
\hline\noalign{\smallskip}
$M_{\rm i}/M_{\odot}  $ &3.50  &3.50  &3.50  &3.30  &3.00  &2.30 \\
\hline\noalign{\smallskip} %\noalign{\smallskip}\hline
\end{tabular}\end{center}
\end{table*}

\subsection{comparison between observations and theoretical predictions }\label{subs:3.3}
A comparison between observations and our theoretical predictions
about IFMRs is shown in Fig. \ref{Fig4}. Some empirical relations
are also presented in the figure. The cross represents the mean
error of the observational data. The solid line is the IFMR for
stars of $Z=0.02$ obtained in this paper. The dashed, dot-dashed
and thick dotted lines are the empirical relations given by
Weidemann (\cite{WEI00}), Ferrario et al. (\cite{Fer05}) and
Williams (\cite{WIL06}), respectively. It is obvious that our
theoretical IFMR for stars of $Z=0.02$ is consistent with the
observations and the empirical relations, especially for low mass
stars. This is a natural result since almost all clusters observed
are open clusters in the Galaxy and their metallicities are around
$0.02$.

\subsection{the potential evidence of the dependence of the initial-final mass relationship on the metallicities}\label{subs:3.4}
The IFMR derived from NGC 2099 might be evidence of the dependence
of the IFMR on the metallicity. Kalirai et al. (\cite{KAL05})
noticed that half of their data points reside in a region of the
IFMRs for stars of $Z=0.008$ and $Z=0.02$ given by Marigo
(\cite{MAR01}). Then, they suggested that they would discover the
first evidence of the effect of the metallicity on the IFMR. We
rehandle the data in Kalirai et al. (\cite{KAL05}). The result is
shown in Fig \ref{Fig5}. The solid line and the dot-dashed line
are the results in this paper for $Z=0.02$ and $Z=0.01$,
respectively. The dashed line is the best-fit linear least squares
line from the data in  Kalirai et al. (\cite{KAL05}). For
convenience to compare with the results in this paper and with the
empirical relations, we set the slope of the fitted line as that
of the empirical relation given by Williams (\cite{WIL06}), which
is consistent with our IFMR for stars of $Z=0.02$ (see the linear
thick dotted line in Fig \ref{Fig4}). The method used here is
similar to that in calculating a mean value since the mean value
is the intercept of a line whose slope is zero. When fitting the
linear line, we use the errors of the data as weight. Seen from
Fig \ref{Fig5}, the fitted line is more consistent with the line
of $Z=0.01$ than that of $Z=0.02$. Comparing with the empirical
relation given by Williams (\cite{WIL06}), the fitted line moves
upward by $0.05 M_{\odot}$. Interestingly, $0.05  M_{\odot}$ is
equal to the estimate of the $\Delta M_{\rm f}$ between the IFMR
for stars of $Z=0.02$ and that of $Z=0.01$ by Williams
(\cite{WIL06}). As shown in subsection \ref{subs:3.3}, the IFMR
for stars of $Z=0.02$ in this paper is consistent with the
observations in the Galaxy. Then, although the observational error
is large, we reconfirm the discovery of Kalirai et al.
(\cite{KAL05}) that the IFMR derived from observation in NGC 2099
is potential evidence of the dependence of the IFMR on the
metallicities and that the metallicity of NGC 2099 may be about
$0.01$. Note that the observational metallicity of NGC 2099 is
[Fe/H]$\sim$-0.1, which is less than solar metallicity (Kalirai et
al. \cite{KAL05}).

Another source of potential evidence is the Hyades. The
metallicity of the Hyades ([Fe/H]=+0.17) is larger than that of
NGC 2099, while its age is similar to that of NGC 2099 (Perryman
et al. \cite{PER98}; Kalirai et al. \cite{KAL01}, \cite{KAL05}).
The mean mass of WD in NGC 2099 is $<M>=0.80\pm 0.03 M_{\odot}$
whereas WDs in the Hyades have a mean mass of $<M>=0.72\pm 0.02
M_{\odot}$ (Claver et al. \cite{CLA01}; Kalirai et al.
\cite{KAL05}). This also qualitatively matches with the trend
obtain in this paper that for $Z<0.04$ -- low metallicity, on
average, leads to higher final mass.

Recently, Kalirai et al (\cite{KAL07b}) showed the first
constraints on IFMRs at the low mass end. Their results are
derived from three different clusters with different ages and
different metallicities, i.e. NGC 7789, NGC 6819 and NGC 6791.
Since the ages and the initial masses for these clusters are
different, their results can not give a direct constraint on
whether the IFMR depends on metallicity. However, our results are
consistent with that of Kalirai et al (\cite{KAL07b}) within the
errors. For example, WD7 in NGC 6791 ($\rm [Fe/H]=+0.4\pm0.1]$)
has $M_{\rm i}=1.16_{-0.03}^{+0.04} M_{\odot}$ and $M_{\rm
f}=0.53\pm 0.02 M_{\odot}$, which is similar to our results of
$Z=0.04$, i.e. $M_{\rm f}= 0.5468 M_{\odot}$ for $M_{\rm i}=1.01
M_{\odot}$ and $M_{\rm f}= 0.5560 M_{\odot}$ for $M_{\rm i}=1.31
M_{\odot}$. The cases of the other two cluster are similar. So,
Kalirai et al (\cite{KAL07b}) might provide indirect evidence of
the dependence of the IFMRs on metallicity.

   \begin{figure}
   \centering
%   \vspace{2mm}
%   \begin{center}
   %%%\includegraphics{empty.eps}
   %%%\includegraphics{empty.eps}
   \includegraphics[width=60mm,height=80mm,angle=270.0]{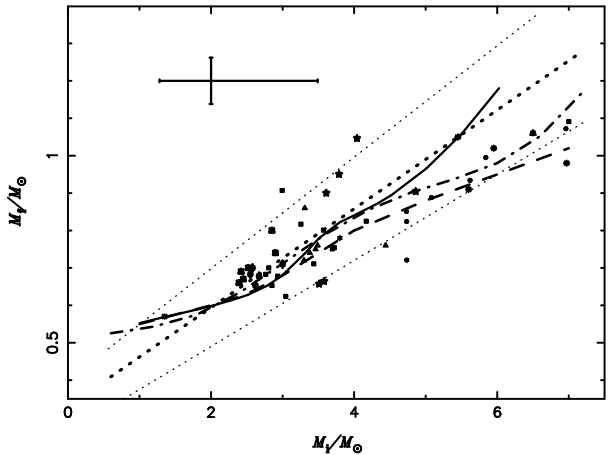}
%   \hspace{3mm}
%   \psfig {figure=mix.ps,width=80mm,height=60mm,angle=270.0}
%   \parbox{180mm}{{\vspace{2mm} }}
   \caption{Comparison between our theoretical prediction and observations.
   The cross represents the mean error of the observational value. Most of the observational data are from the observation
   of some open clusters
   in the Galaxy and the data are from
   Reid (\cite{REI96}), Herwig (\cite{HER95}), Koester \& Reimers (\cite{KR96}), Finley \& Koester (\cite{FK97}),
   Claver et al. (\cite{CLA01}), Williams et al. (\cite{WIL04}), Ferrario et al.(\cite{Fer05}) and Dobbie (\cite{DOB06b}).
   The solid line is the IFMR for stars of  $Z=0.02$ in the paper.
   The dashed, dot-dashed and thick dotted lines are the empirical relations given by Weidemann (\cite{WEI00}),
   Ferrario et al. (\cite{Fer05}) and Williams (\cite{WIL06}), respectively. The two thin dotted lines show the
range of error of the
   thick dotted line. }
              \label{Fig4}%
    \end{figure}
   \begin{figure}
%   \centering
%   \vspace{2mm}
   \begin{center}

   \includegraphics[width=60mm,height=80mm,angle=270.0]{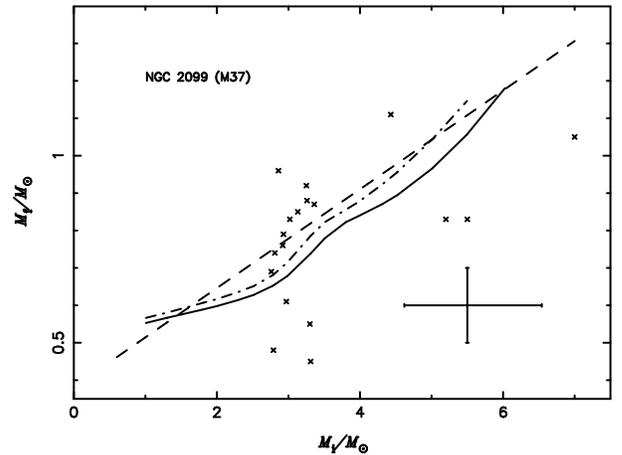}
%   \hspace{3mm}
%   \psfig {figure=mix.ps,width=80mm,height=60mm,angle=270.0}
%   \parbox{180mm}{{\vspace{2mm} }}
   \caption{Comparison between our results and the observation in NGC 2099 (M37). The cross represents
   the mean error of observational data. The solid line and the dot-dashed one are the results in this paper for
   Z=0.02 and Z=0.01, respectively and the dashed one is the best-fit linear
least squares line of the observational data. The data are from
Kalirai et al. ({\cite{KAL05}}).
   }
              \label{Fig5}%
      \end{center}
    \end{figure}

   \begin{figure}
%   \centering
%   \vspace{2mm}
   \begin{center}
   \includegraphics[width=60mm,height=80mm,angle=270.0]{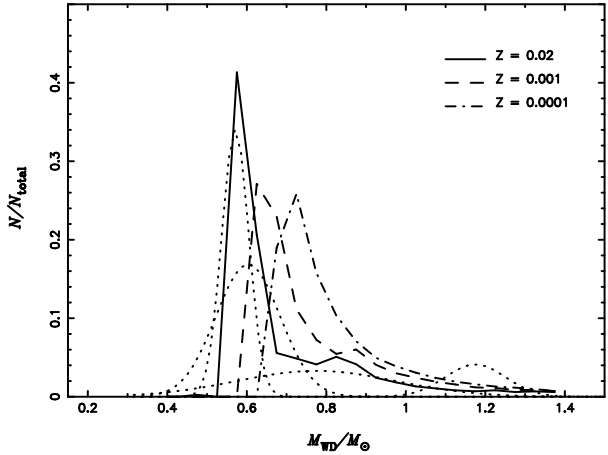}
%   \hspace{3mm}
%   \psfig {figure=mix.ps,width=80mm,height=60mm,angle=270.0}
%   \parbox{180mm}{{\vspace{2mm} }}
   \caption{ Mass distribution of white dwarfs calculated by Hurley's single stellar population synthesis code
   incorporating our initial-final
   mass relationships for stars of different metallicities. Solid line represents the
   mass distribution of white dwarfs for $Z=0.02$, assuming a constant star
formation rate of $3M_{\rm \odot}$/yr. Dashed and dot-dashed lines
represent the
   mass distribution of white dwarfs for $Z=0.001$ and $Z=0.0001$ with a single star burst, respectively.
   The age when WDs form in this diagram
   is less than 15 Gyr for all the metallicity.
   The dotted lines are the Gaussian fits of the mass distribution of single and
non-magnetic DA stars with
   $T_{\rm eff}\geq 12000$K in Data Release 4 of the Sloan Digital Sky Survey (Kepler et al. \cite{KEP06}).
   }
              \label{Fig6}%
      \end{center}
    \end{figure}

   \begin{figure}
   \centering
%   \vspace{2mm}
%   \begin{center}
   %%%\includegraphics{empty.eps}
   %%%\includegraphics{empty.eps}
   \includegraphics[width=60mm,height=80mm,angle=270.0]{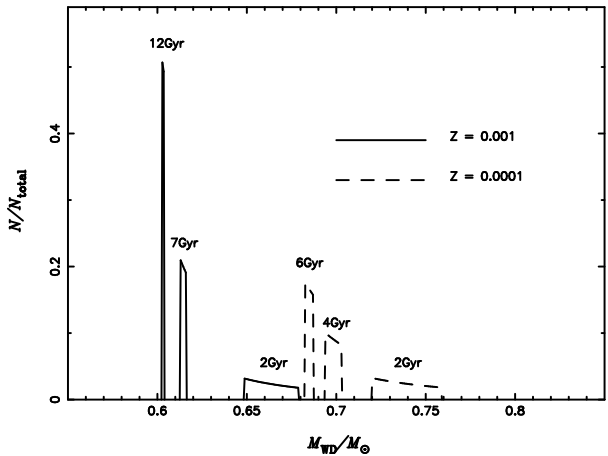}
%   \hspace{3mm}
%   \psfig {figure=mix.ps,width=80mm,height=60mm,angle=270.0}
%   \parbox{180mm}{{\vspace{2mm} }}
   \caption{Mass distribution of white dwarfs with observational selection effect.
   A single starburst is assumed for both metallicities. The age for every line is shown on the top of the line.
   }
              \label{Fig7}%
    \end{figure}

\subsection{the mass distribution of white dwarfs}\label{subs:3.5}
Incorporating the IFMR in this paper into Hurley's single stellar
population synthesis code (Hurley et al. \cite{HUR00},
\cite{HUR02}), we simulate the distribution of the final mass. The
results for age $\leq 15$ Gyr are shown in Fig. \ref{Fig6}. For
$Z=0.02$, a constant star formation rate of $3M_{\rm \odot}$/yr is
assumed to match the observation in the Galaxy (Miller \& Scalo
\cite{MIL79}; Timmes et al. \cite{TIM97}). For $Z=0.001$ and
$Z=0.0001$, a single star burst is assumed. Actually, the star
formation rate does not significantly affect the distribution of
the final mass. Kepler et al. (\cite{KEP06}) showed that the mass
distribution of the single and non-magnetic DA stars in Data
Release 4 of the Sloan Digital Sky Survey (hereafter Sloan DR4)
can be fitted by four Gaussian fits. The Gaussian fits of the mass
distribution of DA stars hotter than $T_{\rm eff}=$12000K from
Sloan DR4 are also shown in Fig. \ref{Fig6} by dotted lines
(Kepler et al. \cite{KEP06}). Incorporating the IFMR for stars of
$Z=0.02$ into Hurley's rapid stellar evolution code, we reproduce
the observational mass distribution of the WDs except for a small
bump at $M_{\rm core}\simeq 1.20 M_{\odot}$. The BPS results of
Han (\cite{HAN98}) showed that the final mass of the merger of two
CO WD concentrates in $M_{\rm f}\simeq 1.20 M_{\odot}$ if the mass
of the merger is less than the Chandrasekha mass limit (see in
subsection \ref{subs:4.2} for details). The bump in Sloan DR4 may
be derived from the mergers of double degenerate systems.

In Sloan DR4, both DA WDs and DB WDs show a trend that the mean
mass decreases with the effective temperature (Kepler et al.
\cite{KEP06}). Kepler et al. (\cite{KEP06}) suggested that this
trend would not be a real case and it would be derived from the
absence of neutral particles in their model used to fit the WD
spectral of Sloan DR4. From our study, however, it is very likely
that the trend is derived from the the effect of the metallicity
as suggested by Wilson (\cite{WIL00}). We know that stars with a
low metallicity evolve faster than those with a high metallicity.
So WDs from stars with a low metallicity form earlier and they
have a longer time to cool. Meanwhile, WDs with a low metallicity
are more massive for a certain initial mass if $Z<0.04$ (as shown
in Figs \ref{Fig1} and \ref{Fig6}), and will cool more quickly
(Kippenhahn \& Weigert \cite{KW90}). Both of these facts imply
that some massive WDs will be cooled and may not be observed. For
the reasons above, we may expect a higher mean mass of WDs in
globular clusters than that in open clusters or in the field.
However, this prediction was not found to hold in NGC 6752, which
is an old low-metallicity globular cluster (Moehler et al.
\cite{MOE04}). The reasons are as follows. Most globular clusters
are very old and high mass WD cools faster than low mass WDs
(Kippenhahn \& Weigert \cite{KW90}). So, the luminosity of high
mass WDs are probably too low to be detected. We construct a toy
model to examine the validity of the explanation. The basic
parameters for the toy model are from Moehler et al.
(\cite{MOE04}), i.e. the distance modulus is set to (m-M)=13$^{\rm
m}$.05-13$^{\rm m}$.20, which is used to deduce the mean value of
the WD mass in the globular cluster (0.53- 0.59 $M_{\odot}$) by
Moehler et al. (\cite{MOE04}); the absolute magnitude of the Sun
is 4$^{\rm m}$.75; the limiting magnitude is set to $26^{\rm m}$
since the signal-to-noise ratio (S/N) is low in Moehler et al.
(\cite{MOE04}) (see their Fig. 1). The lowest luminosity WDs which
can be detected in NGC 6752 is $10^{-3.28}-10^{-3.22} L_{\odot}$,
which means that a WD can not be detected in the cluster NGC 6752
if it cools for about 1Gyr (Fontaine et al. \cite{FON01}; Moroni
\& Straniero \cite{MOR02}, \cite{MOR07}). The mass distributions
of WDs with observational selection effect for different
metallicities and different age are shown in Fig. \ref{Fig7}. The
initial mass function (IMF) is set to $\phi (m)\propto m^{-2.35}$
(Salpeter \cite{SAL55}) and the ages of the progenitors of WDs are
estimated by the life of the progenitors in main sequence
multiplying by a factor of 1.1 (Hurley et al. \cite{HUR00}). The
solid and dashed lines show the mass distribution of WDs for
$Z=0.001$ and $Z=0.0001$, respectively. The age for every line is
shown on the top of the line. It is evident that the position of
the peak in the mass distribution of the WDs moves to a lower mass
with time. If the age of a globular cluster with $Z=0.0001$ is
larger than about 7Gyr, the luminosity of all the WDs with $M_{\rm
i}> 1 M_{\odot}$will be too low to be detected. NGC 6752 is a very
old low-metallicity globular cluster (age is larger than 12Gyr and
$z=0.002$, Cassisi et al. \cite{CAS00}). For the case of
$Z=0.001$, the peak mass of WDs is smaller than about
0.60$M_{\odot}$ at 12Gyr, which is slightly larger than the
observational value, $0.53 M_{\odot} - 0.59 M_{\odot}$.
Considering an observational error of 0.05$M_{\odot}$ and the real
metallicity of NGC 6752, our results match with the observations
of Moehler et al. (\cite{MOE04}). A fact should be noticed that
since NGC 6752 is a globular cluster, the interaction among stars
in the cluster might reduce the final mass.

The results of Fig \ref{Fig7} indicate that the mass distributions
of WDs are possibly a good cosmochronometers as first suggested by
Schmidt (\cite{SCH59}). On the other hand, the mass distribution
of WDs also has a potential ability to distinguish the
metallicitiy of a cluster since the mass peak with a given age is
different for different metallicities. So, the mass distribution
of WDs in globular clusters can help to break the degeneracy of
the metallicity and the age if there are enough WDs to be
observed.

   \begin{figure}
%   \centering
%   \vspace{2mm}
   \begin{center}

   \includegraphics[width=60mm,height=80mm,angle=270.0]{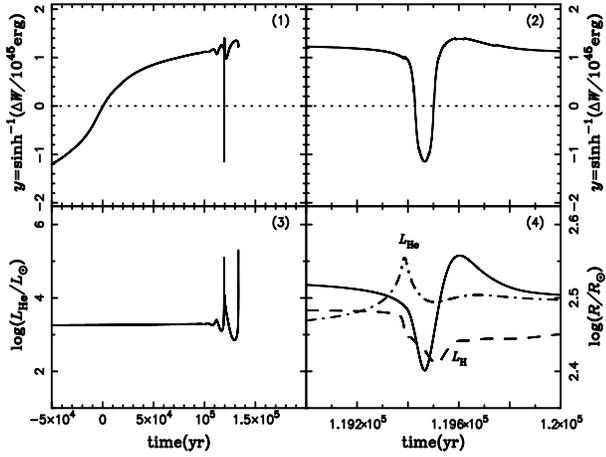}
%   \hspace{3mm}
%   \psfig {figure=mix.ps,width=80mm,height=60mm,angle=270.0}
%   \parbox{180mm}{{\vspace{2mm} }}
   \caption{Binding energy of the envelope (panels 1 and
   2), helium luminosity (panel 3) and stellar radius (panel 4, solid line) as a function of
   time for a star with $M_{\rm i}=3.00 M_{\odot}$
   and $Z=0.02$. The time when $\Delta W=0$
   is set to be zero in the figure. Panels 2 and 4 are the enlarged images of panels 1 and 3 along abscissa.
   The dependent variable of $\Delta W$ in panels 1 and
   2 is the same as that in Fig. \ref{Fig2}. For comparison, helium and hydrogen luminosities
   are also plotted in panel 4 by
   dot-dashed and dashed lines, respectively, and their values are shown by the ordinate of panel 3.
   }
              \label{Fig8}%
      \end{center}
    \end{figure}
\section{Discussions and Conclusions}\label{sect:4}
\subsection{the uncertainty of the initial-final mass relationship }\label{subs:4.1}
The major uncertainty of the IFMR is from the assumption that the
final mass is equal to the core mass when the binding energy of
the envelope $\Delta W=0$ for an AGB/FGB star. Although the
consistency between the mass distribution of WDs in the paper and
that from Sloan DR4 upholds the assumption, there is no direct
observational evidence to verify it. Another widely accepted
choice of the final mass is the core mass of a star at the
beginning of the thermal-pulse AGB (TPAGB, the first TP or the end
of early AGB) (see the review by Weidemann \cite{WEI00}). From
this choice, taking $Z=0.02$ as an example, the general trend of
the IFMR is divided into three segments as follows: (i), the core
mass is almost constant from 1 $M_{\odot}$ to 2.5 $M_{\odot}$,
around 0.52 $M_{\odot}$, which is smaller than that in this paper
by about 0.03-0.1 $M_{\odot}$. (ii), the core mass strongly
increases with the initial mass up to 4.0 $M_{\odot}$. This trend
is similar to that in this paper, but the core mass is smaller
than that in this paper by 0.03-0.05 $M_{\odot}$. (iii), the core
mass slowly increases with the initial mass up to 6-7 $M_{\odot}$.
The final mass is smaller than that in this paper by 0.1-0.2
$M_{\odot}$. According to above discussion , the maximum variation
of the final mass derived from different assumptions may be as
large as 0.2 $M_{\odot}$.

In this paper, we skipped the detailed evolution of TP and simply
treated it as an average evolution, which may lose some
information on the structural change of the envelope caused by
thermal pulse at the early phase of thermally pulsing AGB. We
therefore chose a model with $M_{\rm i}=3.00 M_{\odot}$ and
$Z=0.02$ to examine the influence of thermal pulses on $\Delta W$
and the result is shown in Fig. \ref{Fig8}. From the figure, we
see that there is a time that $\Delta W$ changes its sign but
returns back immediately during a full amplitude of TP. The change
of the sign of $\Delta W$ is from the decrease of stellar radius
which results from the decrease of total luminosity (see panels 3
and 4). This implies that superwind might be periodically
interrupted, but the timescale of the interruption is very short
($\sim100$yr), and then its influence is small. There are many
studies focusing on the TPs previously and we briefly summarize
them in the following. Generally, the final mass after several TPs
is larger than the core mass at the first TP. Forestini \&
Charbonnel (\cite{FC97}) calculated a model of $M_{\rm i}=3
M_{\odot}$ assuming a Reimers mass-loss law with $\eta$ increasing
from 2.5 to 5.0 in the final AGB stage. They obtained $M_{\rm
f}=0.60M_{\odot}$ after 19 TPs, but the core mass is
0.54$M_{\odot}$ at the first TP. Dominguez (\cite{DOM99}) showed
that the thermal pulse starts when the core mass is
0.571$M_{\odot}$ for a star of $M_{\rm i}=3 M_{\odot}$, while the
final mass increases to $M_{\rm f}=0.71 M_{\odot}$ after 26 TPs if
the mass loss prescription of Groenewegen \& de Jong
(\cite{GRO94}) was adopted (Straniero et al. \cite{STRA97}). If
the fairly strong mass-loss law of Bl$\ddot{\rm o}$cker
(\cite{BLOC95}) was adopted, the final mass becomes
0.63$M_{\odot}$ after 20 TPs from 0.53$M_{\odot}$ at the first TP
(Bl$\ddot{\rm o}$cker \cite{BLOC95}). However, the final mass of
0.68 $M_{\odot}$ is obtained from the model with $M_{\rm i}=3.0
M_{\odot}$ and $Z=0.02$ in this paper.

{In Fig. \ref{Fig8}, we see that $\Delta W=0$ for a star with
$M_{\rm i}=3.00 M_{\odot}$ and $Z=0.02$ is achieved before TPAGB
phase, which might be very important for galactic chemical
evolution since it is widely believed that s-process elements are
created in TPAGB phase of a low-mass star (Han et al.
\cite{HAN95}; Busso et al. \cite{BGW99}). Actually, $\Delta W=0$
for all the models in this paper is achieved at EAGB phase. These
results seem to indicate that the s-process elements could not be
created. In fact, as mentioned in subsection \ref{subs:2.1}, it is
more likely that $\Delta W=0$ is only a lower time limit for
superwind, i.e. a superwind starts at or soon after $\Delta W=0$.
Therefore, the stars achieving $\Delta W=0$ at EAGB phase still
have opportunities to enter into TPAGB phase, and contribute to
s-process elements.} So, the final mass shown in this paper might
be different from a real one. As mentioned above, the core mass in
3 $M_{\rm \odot}$ models grows by about 0.1 $M_{\rm \odot}$ during
TPAGB evolution. For a 1.5 $M_{\rm \odot}$ star with $Z=0.02$, the
final core mass is about 0.03 $M_{\rm \odot}$ larger than that in
this paper if the mass loss prescription of Groenewegen \& de Jong
(\cite{GRO94}) was adopted (Dominguez \cite{DOM99}). The main
result in this paper then still holds.

In Fig. \ref{Fig4}, we see that when $M_{\rm i}<3.5M_{\odot}$, the
IFMR in this paper is well consistent with the empirical
relations. However, for the higher initial mass, the IFMR in this
paper is larger than the empirical relations given by Weidemann
(\cite{WEI00}) and Ferrario et al. (\cite{Fer05}). This
discrepancy is mainly from our assumption. As mentioned in
subsection \ref{subs:2.1} and above paragraph, $\Delta W\geq0$ is
a necessary condition to eject the envelope of an AGB star. Since
$\Delta W=0$ is achieved at the second dredge-up for the star with
$M_{\rm i}>3.5M_{\odot}$ during EAGB phase and the core is
decreasing at this phase, the necessary condition means that final
mass may be overestimated. The uncertainty of $M_{\rm f}$
attributed to the definition of core in this paper may also reduce
the discrepancy. Meanwhile, as mentioned in subsection
\ref{subs:3.1}, the final mass might be overestimated if the
helium envelope of a star reaches $\Delta W=0$ again after the
ejection of hydrogen-rich envelope. This overestimate is more
likely for high initial mass, since the helium envelope is
thicker, and then reaches $\Delta W=0$ more easily (see subsection
\ref{subs:3.1}). However, since the errors from the observations
are also very large (see the error bar in Fig. \ref{Fig4}), the
IFMR obtained in this paper is well located in the error range and
therefore, is still consistent with observations.

\subsection{low-mass white dwarf}\label{subs:4.2}
Kalirai et al. (\cite{KAL07a}) suggested that helium white dwarfs
may be obtained in metal-rich old clusters, i.e. NGC 6791. It is a
reasonable assumption that the same mechanism in NGC 6791 would
work for metal-rich field stars. Recently, Kilic, Stanek \&
Pinsonneault (\cite{KIL07c}) rehandled the data in Valenti \&
Fischer (\cite{VF05}) and found that there have been metal-rich
stars with [Fe/H]$>$ 0 at all times in the local Galactic disk,
although the metallicity distribution of disk stars peaks below
solar metallicity for stars with ages greater than about 5 Gyrs.
Considering that only 5\% of all WDs in the local disk are single
low-mass white dwarfs ($<$ 0.45$M_{\odot}$) (Liebert et al.
\cite{LIB05}; Kepler et al. \cite{KEP06}) and the fraction of
metal-rich stars with ages greater than 9 Gyrs is 21\%, Kilic,
Stanek \& Pinsonneault (\cite{KIL07c}) argued that only the stars
of [Fe/H]$>$+0.3 can lose their hydrogen-rich envelope to produce
helium WDs on the FGB. From our study, however, all stars with
$M_{\rm i}\leq1.0 M_{\odot}$ and $Z\geq0.02$ will lose their
hydrogen-rich envelope and finally become helium white dwarfs.
This inconsistency may result from the overestimate of metal-rich
stars in the sample of Valenti-Fischer (Reid et al. \cite{REI07})
adopted by Kilic, Stanek \& Pinsonneault (\cite{KIL07c})

As shown in our study, there should be numerous He WDs in
metal-rich old clusters, which has indeed been observed in NGC
6791 (Kalirai et al. \cite{KAL07a}). Here, we emphasize that we do
not assume any mass-loss mechanism while Kalirai et al.
(\cite{KAL07a}) assumed a metal-enhanced stellar wind on the red
giant branch (RGB). Our study shows that $\Delta W = 0$ in several
FGB stars with high metallicities and small initial masses can be
achieved before these stars get to the tip of FGB, indicting that
the relative number of the tip RGB stars in metal-rich old
clusters is smaller than that in metal-poor clusters. This fact is
directly observed in the RGB luminosity functions of two open
clusters, e.g. NGC 188 and NGC 6791 (see Fig 8 in Kalirai et al
\cite{KAL07a}). Meanwhile, a similar effect should be seen in the
local population of RGB stars. Luck \& Heiter (see Fig 9 in their
paper, \cite{LH07}) make a comparison of the metallicity
histograms between dwarfs and giants within 15 pc of the Sun and
found that the dwarf population has a high metallicity tail
extending up to [Fe/H]$\sim$ 0.6, while the giants show a sharp
drop in numbers after [Fe/H]=0.2 and no giants with [Fe/H]$>$ 0.45
are observed in the field.

The fate that low-mass metal-rich stars will become He WDs before
helium is ignited on FGB might give some constrain on planetary
nebulae (PNe). It is widely accepted that PNe originate from AGB
stars or are related to binary evolution. For the case of binary
evolution, one component in the binary system fills its Roche lobe
at AGB phase and the mass transfer is dynamically unstable which
leads to a common envelope phase. After the ejection of the common
envelope, a PN may form. However, since some metal-rich stars may
not experience AGB phase, we might speculate that the number of
PNe in metal-rich galaxies is relatively lower than that in
metal-poor galaxies. Interestingly, Buzzoni et al (\cite{BUZ06},
see their Fig 11) really observed a relatively low number of PNe
per unit galaxy luminosity in more metal-rich elliptical galaxies.
Gesicki \& Zijlstra (\cite{GZ07}) compared the mass distribution
of the central stars of planetary nebulae (CSPN) with those of
WDs. These two distribution are very different, i.e. the CSPN mass
distribution is sharply peaked at 0.61 $M_{\odot}$ ranging from
0.55 $M_{\odot}$ to 0.66 $M_{\odot}$, while the WD distribution
peaks at a slightly lower mass and shows a much broader range of
masses. Gesicki \& Zijlstra (\cite{GZ07}) suggested that this
difference may imply that only some WDs have gone through the PN
phase. Our models provide a channel for WDs to avoid the PN phase.

Based on our results, WDs from single stars are always larger than
0.4 $M_{\odot}$, whatever the composition of the WD is. Then,
extremely low-mass WDs ($~0.2 M_{\odot}$) are possible in binary
systems or once in binary systems since no stellar population is
old enough to produce such extremely low-mass WDs through single
star evolution. Only recently, several extremely low-mass ($~0.2
M_{\odot}$) WDs were discovered in the field (Liebert et al.
\cite{LIB04}; Kawka et al. \cite{KAW06}; Eisenstein et al.
\cite{EIS06}; Kilic et al. \cite{KIL07a}), and some of them are
companions of pulsars (van Kerkwijk et al. \cite{VANK05}). Kilic
et al. (\cite{KIL07b}) also found a low-mass WD in a binary
system. If an extremely low-mass WD were a single star, it is very
likely that it could have a very high space velocity since it may
come from a close double-degenerate binary, where its companion
has gone through a supernova event that disrupted the binary
(Hansen \cite{HANS03}). LP 400-22 might be a case from this
channel (Kawka et al. \cite{KAW06}). In any case, hard work is
need to systematically search for the companions of WDs with mass
less than 0.4 $M_{\odot}$.

\subsection{the potential effect of the initial-final mass relationship on type Ia supernovae}\label{subs:4.3}
As the best cosmological distance indicators, Type Ia supernovae
(SNe Ia) have been successfully applied to determine the
cosmological parameters ,e.g. $\Omega_{\rm M}$ and $\Omega_{\rm
\Lambda}$ (Riess et al. \cite{REI98}; Perlmutter et al.
\cite{PER99}). Phillips relation (Philips \cite{PHI93}) is used
when taking SNe Ia as the distance indicators. It is assumed that
Phillips relation is correct at high redshift, although the
relation was obtained from a low-redshift sample. This assumption
is precarious since the exact nature of SNe Ia is still unclear,
especially the progenitor model and explosion mechanism
(Hillebrandt \& Niemeyer \cite{HN00}; Leibundgut \cite{LEI00}). If
the properties of SNe Ia evolve with redshift, the results for
cosmology might be different. Since metallicity decreases with
redshift, a good way to study the correlation between the
properties of SN Ia and redshift is to study the correlation
between the properties of SN Ia and metallicity. It is widely
believed that SNe Ia are from thermonuclear runaway of a
carbon-oxygen white dwarf (CO WD) in a binary system. The CO WD
accretes material from its companion to increase its mass. When
its mass reaches its maximum stable mass, it explodes as a
thermonuclear runaway and almost half of the WD mass is converted
into radioactive nickel-56 (Branch \cite{BRA04}). The mass of
nickel-56 determines the maximum luminosity of SNe Ia. The higher
the mass of nickel-56 is, the higher the maximum luminosity is
(Arnett \cite{ARN82}). Some numerical and synthetical results
showed that metallcity may affect the final amount of nickel-56,
and thus the maximum luminosity of SNe Ia (Timmes et al.
\cite{TIM03}; Travaglio et al. \cite{TRA05}; Podsiadlowski et al.
\cite{POD06}). There is also much evidence about the correlation
between the properties of SNe Ia and metallicity in observations
(Branch \& Bergh \cite{BB93}; Hamuy et al. \cite{HAM96}; Wang et
al. \cite{WAN97}; Cappellaro et al. \cite{CAP97}), e.g. the
maximum luminosity of SNE Ia is proportional to the metallicity
(Shanks et al. \cite{SHA02}).

Two progenitor models of SNe Ia have competed for about three
decades. One is a single degenerate model, which is widely
accepted (Whelan \& Iben \cite{WI73}). In this model, a CO WD
increases its mass by accreting hydrogen- or helium-rich matter
from its companion, and explodes when its mass approaches the
Chandrasekhar mass limit. The companion may be a main-sequence
star (WD+MS) or a red-giant star (WD+RG) (Yungelson et al.
\cite{YUN95}; Li et al. \cite{LI97}; Hachisu et al. \cite{HAC99a},
\cite{HAC99b}; Nomoto et al. \cite{NOM99}; Langer et al.
\cite{LAN00}). Hachisu \& Kato (\cite{HK03a}, \cite{HK03b})
suggested that supersoft X-ray sources, which belong to WD+MS
channel, may be good candidates for the progenitors of SNe Ia. The
discovery of the companion of Tycho's supernova also verified the
reliability of the model (Ruiz-Lapuente et al. \cite{RUI04}; Ihara
et al. \cite{IHA07}). Another progenitor model of SNe Ia is a
double degenerate model (Iben \& Tutukov \cite{IBE84}; Webbink
\cite{WEB84}), in which a system consisting of two CO WDs loses
orbital angular momentum by gravitational wave radiation and
merges. The merger may explode if the total mass of the system
exceeds the Chandrasekhar mass limit (see the reviews by
Hillebrandt \& Niemeyer \cite{HN00} and Leibundgut \cite{LEI00}).
In both of the progenitor models, the CO WD which finally explodes
as a SN Ia should approach or exceed the Chandrasekhar mass limit.
Obviously, a higher mass CO WD may fulfill this situation more
easily than a low mass one. According to our results, the mass of
a CO WD with a given initial mass will be higher in the
circumstance with extremely high metallicity or extremely low
metallicity than that in the middle-metallicity circumstance.
Metallicity is therefore very relevant to SNe Ia.

\subsection{Conclusions}\label{subs:4.3}
We use the method of Han et al. (\cite{HAN94}) to calculate the
IFMR for stars of different metallicities. The conclusions are as
follows.
\begin{enumerate}
\item There is an obvious dependence of the IFMRs on the
metallicity and the dependence is not monotone. When $Z=0.04$, the
final mass of the CO WD for a given initial mass is smallest. For
higher or lower $Z$, the mass of CO WD will be higher. The
difference of the final mass derived from different metallicties
is up to $0.4 M_{\odot}$.

\item The initial-final mass relationship for stars of $Z=0.02$ is
consistent with observations.

\item For $Z\geq0.02$, a helium white dwarf is formed from a star
of $M_{\rm i}\leq 1.0M_{\odot}$. The final masses of helium WDs
for a given initial mass slightly decreases with metallicity.

\item Incorporating the IFMR for stars of $Z=0.02$ in the paper
into Hurley's single stellar population synthesis code, we
reproduce the mass distribution of DA WDs in Sloan DR4 except for
some extra-massive WDs.

\item We reconfirm the discovery of Kalirai et al. (\cite{KAL05})
that the initial-final mass relationship derived from observation
in NGC 2099 might be evidence of the dependence of the IFMR on
metallicities and that the metallicity of NGC 2099 may be about
$0.01$, although the observational error of white dwarfs in NGC
2099 is large. It should be encouraged to program more accurate
observations to find a larger sample of white dwarfs in globular
clusters. Such programs may help to confirm the dependence of the
initial-final mass relationship on the metallicity.

\item We bring up again Willson's suggestion (Willson
\cite{WIL00}) that the effect of metallicity may be the origin of
the phenomenon that the mean mass of WDs decrease with effective
temperature.

\end{enumerate}

\begin{acknowledgements}
We thank Dr Richard Pokorny for his kind help to make this paper
more beautiful. This work was in part supported by the Natural
Science Foundation of China under Grant Nos. 10433030, 10521001,
10603013 and 2007CB815406 and Chinese Academy of Science under
Grant No. O6YQ011001, Yunnan National Science Foundation (No.
2004A0022Q).
\end{acknowledgements}

\newpage
\appendix
%\hspace{0.5mm}\\
\section[]{FITTED FORMULAE OF INITIAL-FINAL MASS RELATIONSHIP FOR STARS of DIFFERENT METALLICITIES}

The initial-final mass relationship for stars of different
metallicities can be approximated by two parabola segments except
for the case of $Z=0.0001$. In this Appendix, we present the
fitted formulae. Hereafter, $M_{\rm i}$ is the initial mass of
stars and $M_{\rm f}$ is the final mass of stars.

For $Z=0.1$,
\begin{equation}
M_{\rm f}=\max [M_{\rm f,1},M_{\rm f,2}],    1.0 M_{\odot}< M_{\rm
i}\leq 5.5 M_{\odot}, {\rm err}<1.0\%,
\end{equation}
where
\begin{equation}
M_{\rm f,1}=0.5675+0.1027M_{\rm i}-0.0006261M_{\rm i}^{\rm 2},
\end{equation}
and
\begin{equation}
M_{\rm f,2}=0.7345-0.07362M_{\rm i}+0.03973M_{\rm i}^{\rm 2}.
\end{equation}

For $Z=0.08$,
\begin{equation}
M_{\rm f}=\min [M_{\rm f,1},M_{\rm f,2}], 1.0 M_{\odot}\leq M_{\rm
i}\leq 6.0 M_{\odot}, {\rm err}<1.8\%,
\end{equation}
where
\begin{equation}
M_{\rm f,1}=0.7097-0.1930M_{\rm i}+0.08235M_{\rm i}^{\rm 2},
\end{equation}
and
\begin{equation}
M_{\rm f,2}=0.8316-0.1183M_{\rm i}+0.03511M_{\rm i}^{\rm 2}.
\end{equation}

For $Z=0.06$,
\begin{equation}
M_{\rm f}=\min [M_{\rm f,1},M_{\rm f,2}], 0.8 M_{\odot}\leq M_{\rm
i}\leq 6.5 M_{\odot}, {\rm err}<2.3\%,
\end{equation}
where
\begin{equation}
M_{\rm f,1}=0.5806-0.06852M_{\rm i}+0.03928M_{\rm i}^{\rm 2},
\end{equation}
and
\begin{equation}
M_{\rm f,2}=0.8957-0.1313M_{\rm i}+0.03004M_{\rm i}^{\rm 2}.
\end{equation}

For $Z=0.05$,
\begin{equation}
M_{\rm f}=\min [M_{\rm f,1},M_{\rm f,2}], 0.8 M_{\odot}\leq M_{\rm
i}\leq 6.5 M_{\odot},  {\rm err}<1.8\%,
\end{equation}
where
\begin{equation}
M_{\rm f,1}=0.5736-0.06446M_{\rm i}+0.03621M_{\rm i}^{\rm 2},
\end{equation}
and
\begin{equation}
M_{\rm f,2}=0.8971-0.1255M_{\rm i}+0.02750M_{\rm i}^{\rm 2}.
\end{equation}

For $Z=0.04$,
\begin{equation}
M_{\rm f}=\min [M_{\rm f,1},M_{\rm f,2}], 0.8 M_{\odot}\leq M_{\rm
i}\leq 6.5 M_{\odot}, {\rm err}<1.9\%,
\end{equation}
where
\begin{equation}
M_{\rm f,1}=0.5737-0.06207M_{\rm i}+0.03353M_{\rm i}^{\rm 2},
\end{equation}
and
\begin{equation}
M_{\rm f,2}=0.8691-0.1107M_{\rm i}+0.02491M_{\rm i}^{\rm 2}.
\end{equation}

For $Z=0.03$,
\begin{equation}
M_{\rm f}=\min [M_{\rm f,1},M_{\rm f,2}], 0.8 M_{\odot}\leq M_{\rm
i}\leq 6.5 M_{\odot},{\rm err}<2.3\%,
\end{equation}
where
\begin{equation}
M_{\rm f,1}=0.5701-0.05225M_{\rm i}+0.03013M_{\rm i}^{\rm 2},
\end{equation}
and
\begin{equation}
M_{\rm f,2}=0.9897-0.1608M_{\rm i}+0.03022M_{\rm i}^{\rm 2}.
\end{equation}

For $Z=0.02$,
\begin{equation}
M_{\rm f}=\min [M_{\rm f,1},M_{\rm f,2}], 0.8 M_{\odot}\leq M_{\rm
i}\leq 6.0 M_{\odot}, {\rm err}<2.1\%
\end{equation}
where
\begin{equation}
M_{\rm f,1}=0.5716-0.04633M_{\rm i}+0.02878M_{\rm i}^{\rm 2},
\end{equation}
and
\begin{equation}
M_{\rm f,2}=1.1533-0.2422M_{\rm i}+0.04091M_{\rm i}^{\rm 2}.
\end{equation}

For $Z=0.01$,
\begin{equation}
M_{\rm f}=\min [M_{\rm f,1},M_{\rm f,2}], 0.8 M_{\odot}\leq M_{\rm
i}\leq 5.5 M_{\odot},{\rm err}<2.2\%
\end{equation}
where
\begin{equation}
M_{\rm f,1}=0.5897-0.05631M_{\rm i}+0.03395M_{\rm i}^{\rm 2},
\end{equation}
and
\begin{equation}
M_{\rm f,2}=0.8660-0.1240M_{\rm i}+0.03183M_{\rm i}^{\rm 2}.
\end{equation}

For $Z=0.004$,
\begin{equation}
M_{\rm f}=\min [M_{\rm f,1},M_{\rm f,2}], 0.8 M_{\odot}\leq M_{\rm
i}\leq 5.0 M_{\odot},{\rm err}<2.2\%,
\end{equation}
where
\begin{equation}
M_{\rm f,1}=0.5983-0.04908M_{\rm i}+0.03780M_{\rm i}^{\rm 2},
\end{equation}
and
\begin{equation}
M_{\rm f,2}=0.8529-0.1056M_{\rm i}+0.03194M_{\rm i}^{\rm 2}.
\end{equation}

For $Z=0.001$,
\begin{equation}
M_{\rm f}=\min [M_{\rm f,1},M_{\rm f,2}], 0.8 M_{\odot}\leq M_{\rm
i}\leq 4.5 M_{\odot},{\rm err}<1.0\%,
\end{equation}
where
\begin{equation}
M_{\rm f,1}=0.5959-0.007419M_{\rm i}+0.03297M_{\rm i}^{\rm 2},
\end{equation}
and
\begin{equation}
M_{\rm f,2}=0.6312-0.04458M_{\rm i}+0.04213M_{\rm i}^{\rm 2}.
\end{equation}

For $Z=0.0003$,
\begin{equation}
M_{\rm f}=\min [M_{\rm f,1},M_{\rm f,2}], 1.0 M_{\odot}\leq M_{\rm
i}\leq 4.5 M_{\odot},{\rm err}<0.4\%,
\end{equation}
where
\begin{equation}
M_{\rm f,1}=0.6300+0.007371M_{\rm i}+0.02700M_{\rm i}^{\rm 2},
\end{equation}
and
\begin{equation}
M_{\rm f,2}=0.7629-0.03288M_{\rm i}+0.02390M_{\rm i}^{\rm 2}.
\end{equation}

For $Z=0.0001$,
\begin{equation}
%\[
% \left
\begin{array}{ll}
M_{\rm f}=0.6033+0.06839M_{\rm i}+0.008401M_{\rm i}^{\rm 2}, \\
 1.0M_{\odot}\leq M_{\rm i}\leq 4.5 M_{\odot}, {\rm err}<1.0\%.
\end{array}
 %\right
%\]
\end{equation}
\end{document}